\documentstyle[12pt,psfig]{article} 
\def\Li#1{{\rm Li}_{#1}(x)}
\newcommand{\be}{\begin{eqnarray}}
\newcommand{\ee}{\end{eqnarray}}

\newcommand{\op}{\hat O}

\newcommand{\ew}{\epsilon _W}
\newcommand{\ez}{\epsilon _Z}
\newcommand{\gu}{{g_{-}^{u}}}
\newcommand{\gd}{{g_{-}^{d}}}
\newcommand{\wm}{L}

\begin{document}

\rightline{TTP96-33}
\rightline{hep-ph/9609302}
\rightline{September 1996}
\bigskip
\begin{center}
{\Large {\bf An estimate of the QCD corrections to the decay}}\\[2mm]
{\Large{\bf \boldmath$Z \to Wu\bar d$}}
\end{center}
\vspace{0.2cm}
\smallskip
\begin{center}
{\large{\sc Andrzej Czarnecki}}\footnote{E-mail:
                 {\tt ac@ttpux8.physik.uni-karlsruhe.de}}\\
{\sl Institut f\"{u}r Theoretische Teilchenphysik}\\
{\sl Universit\"{a}t Karlsruhe}\\
{\sl D--76128 Karlsruhe, Germany}\\
\vspace{0.7cm}
and\\
\vspace{0.7cm}
{\large{\sc Kirill~Melnikov}}\footnote{E-mail:
                {\tt melnikov@dipmza.physik.uni-mainz.de}
  \\ \hspace*{4.65mm} Address after Nov.~1st, 1996: 
    Institut f\"{u}r Theoretische Teilchenphysik,
    Universit\"{a}t Karlsruhe,
    D--76128 Karlsruhe, Germany
          }\\
{\sl Institut f\"ur Physik}\\
{\sl Johannes Gutenberg Universit\"{a}t}\\
{\sl D-55099 Mainz, Germany}\\
\vspace{1.8cm}
{\large{\bf Abstract}}\\
\end{center}
\vspace{0.1cm}
We present an estimate of perturbative 
QCD corrections to the decay  $Z \to Wu\bar d$. 
A simple approximate approach is described in detail.  The difference
of masses of $M_Z$ and $M_W$ is used as an expansion parameter.  
A complete analytical formula for a part of the corrections is also
presented. 

\newpage

\section{ Introduction}
The decay channel $Z \to W X$ was suggested in the early days
of the Standard Model (SM) as a possible source of the $W$--bosons
\cite{MW}. A number of detailed studies arrived at the conclusion that
the branching ratio of this decay mode is extremely small. There are several 
factors which contribute to this strong suppression. First, it is a higher
order electroweak process; second, it is suppressed due to the
small difference in masses of the $Z$ and $W$ bosons; and third, it
suffers from a destructive interference among contributions 
of different diagrams.  On the other hand it was also pointed out that
such a decay process would lead to an exceptional signature (in the
case when W boson decays leptonically). Namely, there will be one charged 
lepton with the transverse energy $E_{\bot} \sim 40$ GeV accompanied
by the same amount of missing energy. If the $W$ decays hadronically
it will be more difficult to isolate the signal from the background.

\vspace*{10mm}
\begin{figure}[htb]
\hspace*{-5mm}
\begin{minipage}{16.cm}
\[
\mbox{
\hspace*{10mm}
\begin{tabular}{ccc}
\psfig{figure=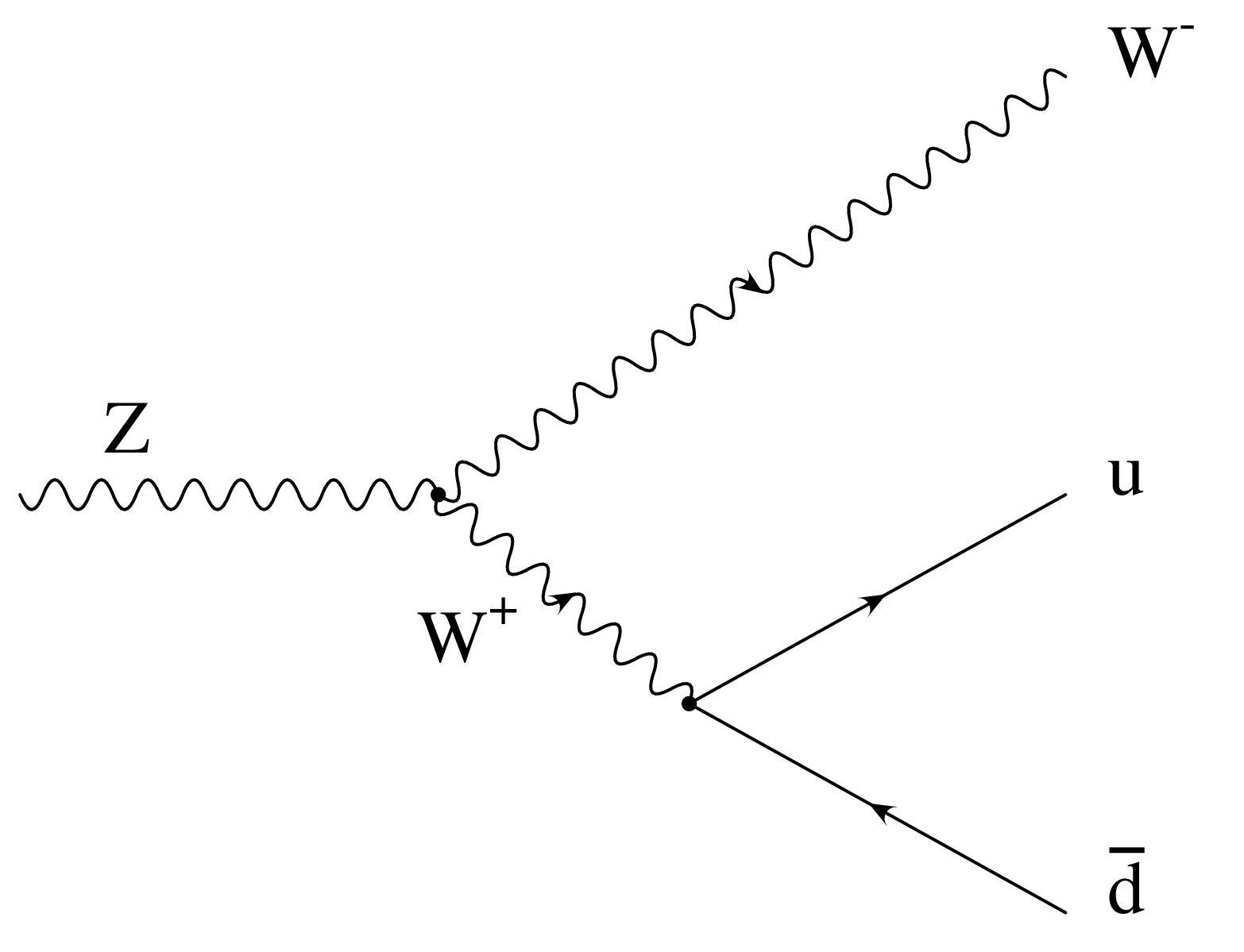,width=27mm,bbllx=210pt,bblly=410pt,%
bburx=630pt,bbury=550pt} 
&\hspace*{15mm}
\psfig{figure=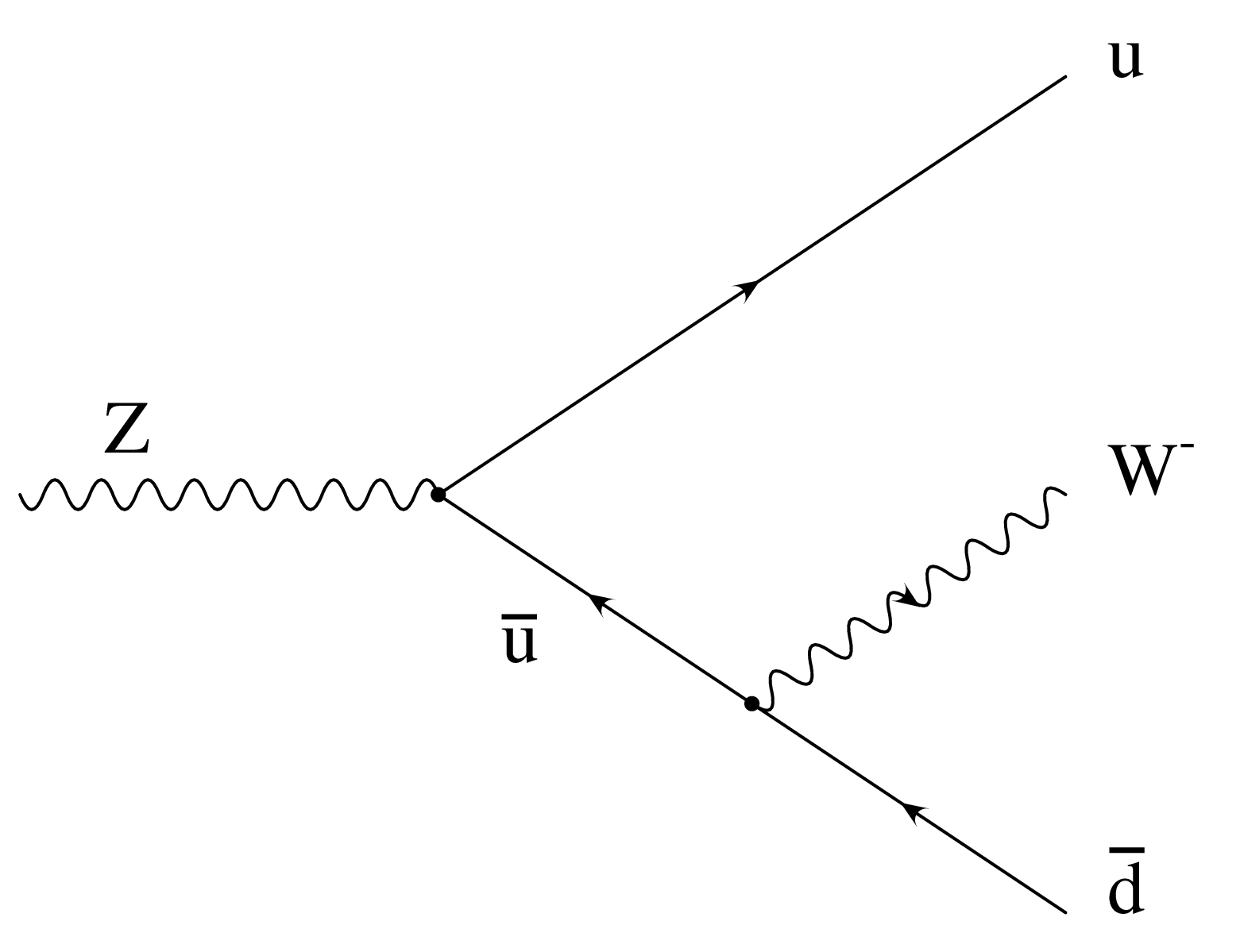,width=27mm,bbllx=210pt,bblly=410pt,%
bburx=630pt,bbury=550pt}
&\hspace*{15mm}
\psfig{figure=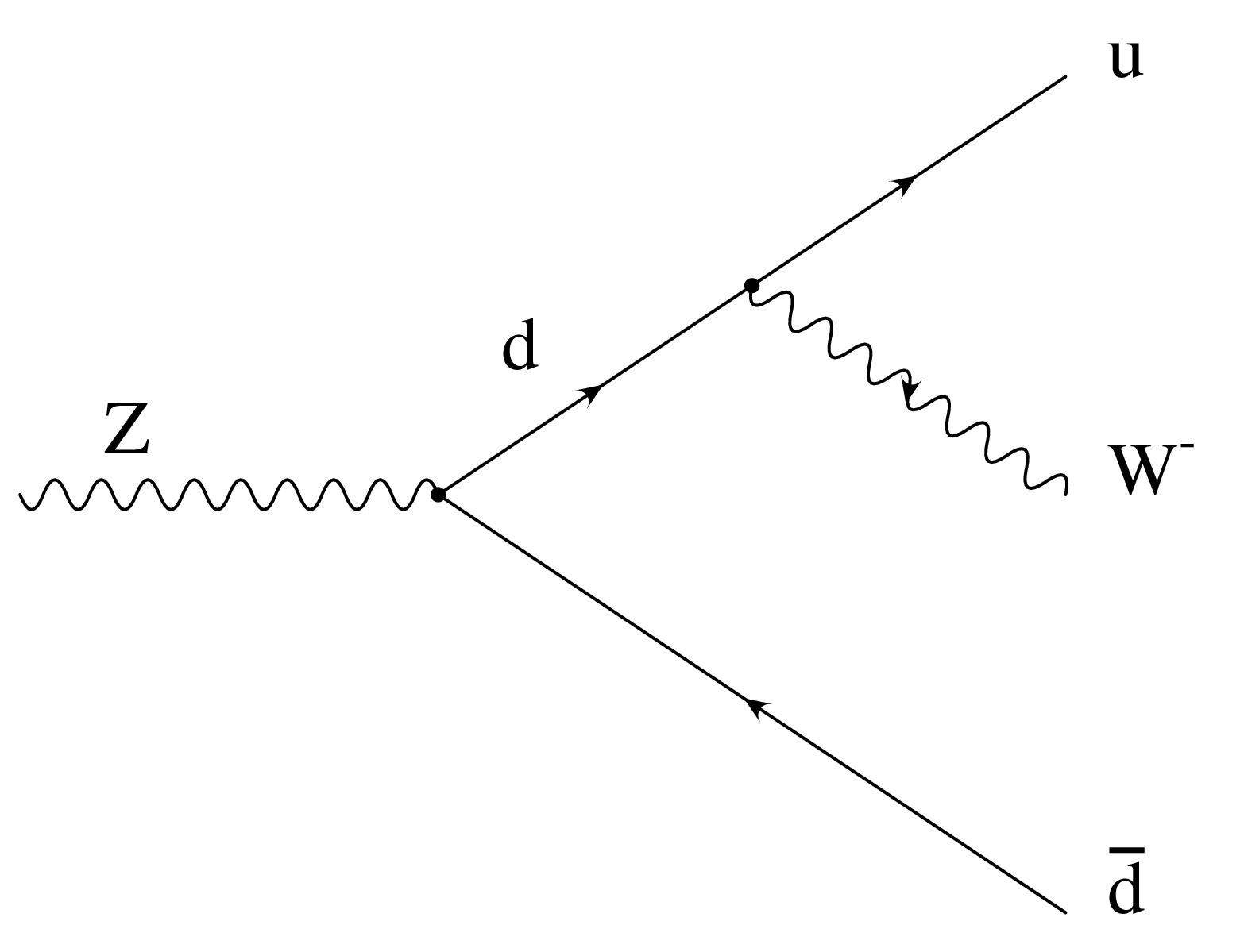,width=27mm,bbllx=210pt,bblly=410pt,%
bburx=630pt,bbury=550pt}
\\[20mm]
(a) & (b) & (c) 
\end{tabular}}
\]
\end{minipage}
\vspace*{10mm}
\caption{Contributions to the decay $Z\to Wu\bar d$ in the Born
  approximation} 
\label{figs}
\end{figure}

An interesting feature of this process is that it involves the 
three gauge boson coupling $ZWW$ (see fig.~1a). 
This coupling gives the largest
contribution to the decay rate. Unfortunately, there are also large gauge 
cancellations from the interference of the diagram with three gauge 
boson coupling with the other diagrams involved. In the SM the coupling
$ZWW$ is fixed. It is likely that effects of physics beyond the SM
can modify this coupling and lead to the deviations from its SM value.
In the context of the $Z \to WX$ decay channel this possibility has been 
analyzed in Ref.~\cite {LS}. It has been concluded that in a special 
case the introduction of anomalous couplings the rate can be increased
by up to one order of magnitude.

Let us present some numbers from Ref.~\cite {Bij}
concerning the number of events one can expect per $10^7$ $Z$ events
in the Standard Model:
$$
N(Z \to Wl\nu \to l\nu l\nu) =0.19,
$$
$$
N(Z \to Wl\nu \to q\bar q \l\nu) =0.38,
$$
$$
N(Z \to Wq\bar q \to l\nu q\bar q) =0.35,
$$
$$
N(Z \to Wq \bar q \to q\bar q q\bar q) =0.70.
$$

Hence to observe the decay $Z \to WX$ within the SM requires more than 
$10^8$ $Z$ events. The largest sample of $Z$ events has been collected by
the four experiments at  LEP1. 
Though the number of $Z$'s is huge ($\sim 16\cdot 10^6$) it is still not 
sufficient to observe the decay $Z \to WX$.  On the other hand, the
designed luminosity of a future $e^+e^-$ collider will allow to
produce $10^7$ $Z$ bosons per day, if the machine operates on the $Z$
resonance.  Such experiments are indeed being planned at the JLC and
will allow a study of the rare decay channels of $Z$ and in particular
of $Z \to WX$.

In Ref.~\cite {BK} the QCD corrections to this decay mode
were discussed. The authors of that paper mainly concentrated on the
non--perturbative and logarithmic perturbative corrections. 
They concluded that the considered corrections are small
and can not strongly shift the lowest order branching ratio. Nevertheless,
pure perturbative corrections ${\cal O}(\alpha _s(M_Z-M_W))$ remained unknown.

In this paper we present a simple estimate of these corrections, based
on the expansion of the rate with respect to the small ratio 
$(M_Z-M_W)/M_Z$. In the next section we demonstrate the basic idea of our
approach with the example of the lowest order decay rate. Then we discuss
the QCD corrections; our results are summarized in
the Conclusions.

\section{Lowest order decay rate}
Let us describe the idea of the calculation with the example of the
Born approximation.  We consider $M_Z$ and $M_W$ as large parameters
(compared to $M_Z-M_W$).  Then the quarks in the final state will have
relatively small energies and momenta. Hence we can try to expand the
amplitude for $Z\to W u\bar d$ in the quantities of the order of
$(M_Z-M_W)/M_Z$.  

We adopt the following notation: $p$ and $k$ are momenta of $Z$
and $W$,  $p_1$ and $p_2$ are momenta of the quarks in the final state.

We begin with the diagram of Fig.~1b
 where the  $W$--boson is emitted from a quark line.
The virtual quark is far off-shell.  Contracting its propagator to a 
point and neglecting momenta of the final quarks we obtain 
the following expression for the amplitude:
\begin{equation}
{\cal M}_2= \frac {ie^2\gu}{\sqrt {2} s_W M_W^2} \bar u(p_1)\op
v(p_2),
\qquad \op =  \hat \epsilon _Z \hat p \hat \epsilon _W \wm .
\label {1}
\end {equation}
Here $\wm = (1-\gamma _5)/2$ and $g_-^q$ describes  $Z$ coupling to
the left-handed quarks, $g_-^q=I_3^q/s_W$.   $s_W$ denotes 
the sine of the weak mixing angle.
We note that according to our approximation we put its cosine equal to
1.

Using Dirac algebra identities we can rewrite this as
\begin{equation}
{\cal M}_2= -~\frac {ie^2\gu}{\sqrt {2} s_W M_W^2}
(C _\mu +A _\mu)
   \bar u(p_1) \gamma ^\mu \wm v (p_2)
\label {2}
\end {equation}
where
\begin{equation}
C _\mu =p_\mu \epsilon _W \cdot \epsilon _Z,~~~
A _\mu =i\epsilon _{\mu \alpha \beta \sigma}
 p ^{\alpha}{\epsilon _W}^{\beta}{\epsilon _Z}^{\sigma}. 
\label {3}
\end{equation}
In deriving this equation we have used the fact that:
\begin{equation}
\epsilon _W\cdot p =\epsilon _Z\cdot p =0,
\label {4}
\end{equation}
which is exactly valid if $M_W=M_Z$.

The interference of 
the two pieces of the amplitudes, proportional to   
$C _\mu$ and $A _\mu$, vanishes after 
the sum over $W$ and $Z$ polarizations.
Note also that the amplitude ${\cal M}_3$ is described by the same equation, 
only the sign of $C_\mu$ is opposite.

The amplitude ${\cal M}_1$ involves a three gauge boson coupling.
Taking there the limit $p_1+p_2\to 0$ we arrive at the approximate formula:
\begin{equation}
{\cal M}_1=\frac {ie^2\sqrt {2}}{s_W^2 M_W^2} (\epsilon _W \cdot \epsilon
   _Z) p _{\mu} 
   \bar u(p_1) \gamma _\mu \wm v (p_2).
\label {5}
\end{equation}

To calculate the leading asymptotics of the Born width we 
integrate over the light quark phase space and then over $W$ three--momenta.
Because of the approximate equality $M_Z \approx M_W$ the $W$ boson is always 
non--relativistic. We approximate its phase space element by
\begin{equation}
\frac {d^3k}{2E_k} \approx \frac {4\pi}{2M_W} k^2dk,
\qquad k_{max}=\frac {M_Z}{2}  x, \qquad x\equiv {M_W^2\over M_Z^2}.
\label {6}
\end{equation}

By counting  the powers of $(M_Z-M_W)$ 
we  see that the leading order
behavior of the tree--level width is given by $(1-x)^5$.

The decay width for $Z \to W u\bar d$ can be written as 
\begin{equation}
\Gamma (Z \to W u \bar d) = \Gamma _0 
 \left[ \frac {\gu ^2 + \gd ^2}{2} 
H_1 (x) + \frac {1}{s_W^2} H_2(x)+\gu \gd H_3(x)-
\frac {(\gu-\gd)}{s_W} H_4(x) \right].
\label {8}
\end{equation}
In the above equation we denote 
\begin{equation}
\Gamma _0 = \frac {N_cM_Z\alpha ^2}{16\pi s_W^2}
\label {8.1}
\end{equation}
and
$H_i(x)$ are the contributions of various diagrams
to the width. 
Namely, $H_1(x)$ is the contribution of the squares of the
graphs $1b$ or $1c$, $H_2(x)$ is the contribution of the
graph $1a$ squared, 
$H_3(x)$ is $1b-1c$ interference and $H_4(x)$ is determined
by $1a-1b$ and $1a-1c$ interferences.

The leading asymptotics of these functions are
\begin{equation}
H_i(x)=(1-x)^5 h_i^{(0)} \Big ( 1 + C_F \frac {\alpha _s}{\pi} \delta _i 
\Big );
\label {8.2}
\end{equation}
$$
h^{(0)}_1 = \frac {1}{20},~~h^{(0)}_2 = \frac {1}{30},
~~h^{(0)}_3 = \frac {1}{60},~~h^{(0)}_4 = \frac {1}{30}.
$$
The results of a  complete Born level 
calculation can be found in \cite {MW}.
Let us note that since we use the values $\gu =-\gd={1\over 2s_W}$
we obtain at Born level complete cancellation between $H_2$ and $H_4$,
potentially the largest contributions to the width. This is the destructive
interference which makes the rate additionally suppressed.

Our  aim in this paper  is to discuss the QCD corrections to this
decay width which are parametrized by the coefficients $\delta_i$.

\section{ QCD corrections} 

First, consider QCD corrections to the square of the amplitude ${\cal M}_1$.
Since the gluons  couple to the external quarks the only  QCD correction
to $|{\cal M}_1|^2$ will be the correction to the correlator of two
$V-A$ currents, 
which can be taken from $W$--decay.
This has been calculated for the first time in Ref.~\cite{Alb}.
The
ratio of the one--loop correction to the Born contribution 
is\footnote{This result is true not only for the 
leading asymptotics in $M_Z-M_W$.} $\alpha _s/\pi$.

Next, we analyze  the QCD corrections to the remaining amplitudes.
As an example let us take the graph were $W$--boson is emitted from the
down quark line (${\cal M}_2$). 
We discuss first the radiation of real gluons.

There are three graphs describing real gluon emission.  In two of them
the gluon is emitted from the external fermion line and in the third
one from the fermion line with high (${\cal O}(M_W^2)$) virtuality.
The energy of the gluon is restricted due to the phase space
suppression, therefore the third graph will have an additional
suppression factor ${\cal O}((M_Z-M_W)^2/M_Z^2)$ relative to the Born
one. It is therefore of no interest for us.

The leading contribution of the two remaining graphs can be found
by contracting the virtual fermion propagator carrying the momentum
${\cal O}(M_{W,Z})$ to a point. 
The effective vertex which appears as the result
of such contraction is equivalent to the effective vertex presented in the
eq.~(\ref {1}). Unfortunately, the transformation which has been used in the
transition from eq.~(\ref {1}) to eq.~(\ref {2}) can not be applied for the 
graphs which describe radiative corrections. The reason for this is the use
of  dimensional regularization and the presence of infra--red 
divergences in the graphs with emission of real gluons.
This problem makes it necessary to perform an ``honest'' calculation
of the ${\cal O}(\alpha _s)$ correction to the production of hadrons by the 
operator $\op$ defined in eq.~(\ref {1}).

Next, it is necessary to analyze virtual corrections.
There are four of them. Note that there are no self energy corrections
to the external quark lines because they are represented by no-scale 
diagrams which vanish in dimensional regularization.  
If we perform on--shell renormalization for the external quark 
lines this  also means that there is no wave--function renormalization
at all.   Therefore the sum of virtual and
real corrections must be finite on its own.

There is a simple way to rearrange virtual contributions to make them
more transparent.  First let us note that in the vertex corrections
to  $Z\to \bar u^{*} u $ or $\bar u^{*} \to W d $ 
we can simply put the momenta 
of the ``soft'' quark equal to zero;  this will not introduce any new 
divergences. The same also applies to the self energy correction to the
virtual fermion line. 

The remaining contribution is due to the box diagram. 
Here the situation is slightly more delicate --- we can not put the momenta 
of the light quarks equal to zero because this leads to
additional divergences. The solution is to apply the ideas of asymptotic 
expansions \cite{ae}. 
In that approach the contribution of the box graph can be obtained 
as a sum of two pieces: in the first one the large momentum flows through 
the fermion propagator which connects the $Z$ and $W$ vertices. To get 
the leading contribution we must contract it to the point. The second piece 
is the expansion of the whole box in Taylor series in the small external 
momenta. In both of these pieces one gets additional divergences 
(ultraviolet in the ``contracted'' part and infrared in the part 
originating from the Taylor series). These divergences are canceled in the
sum. However, it is very useful to separate these pieces because of the
following observation: 
The ``contracted'' part 
of the box   taken together with the real emission graphs will
give a complete ${\cal O}(\alpha _s)$ correction to the production of hadrons
by an operator $\op$ in  eq.~(\ref {1}). Therefore, this sum should be
ultraviolet and infrared finite.
For obvious reasons we will call  this contribution ``soft''.

The remaining (``hard'') virtual corrections, i.e.~corrections to the 
$Z\to \bar u ^* u$ vertex,  $\bar u ^*\to W \bar d $ vertex, self energy 
correction to the $\bar u^*$ propagator, and the box graph, all taken at 
zero momenta of external fermions, must also be  finite. 
Note, however,
that individual pieces are divergent 
and hence one has to perform the Dirac algebra in $D$--dimensions.
The final expression for the hard correction to the amplitude which can be 
obtained in this way is remarkably simple:
\begin{equation}
{\cal M}_{2}^{hard}=~\frac {ie^2\gu}{\sqrt {2} s_W M_W^2}
\frac {C_F\alpha _s}{4\pi} 
(2\,C _\mu +7\,A _\mu)
   \bar u(p_1) \gamma _\mu \wm v (p_2).
\label {9}
\end{equation}

To get an expression for ${\cal M}_{3}^{hard}$ it is sufficient to reverse 
the sign in front of $C_{\mu}$ in eq.~(\ref {9}). One gets
\begin{equation}
{\cal M}_{3}^{hard}=~\frac {ie^2\gu}{\sqrt {2} s_W M_W^2}
\frac {C_F\alpha _s}{4\pi} 
(-~2\,C _\mu +7\,A _\mu)
   \bar u(p_1) \gamma _\mu \wm v (p_2).
\label {9.5}
\end{equation}

We remind that there is no interference between $C_\mu$ and $A_\mu$ 
structures, hence if we know their relative contributions to the Born
width it is a trivial exercise to find a value of the QCD correction.

Let us demonstrate how this works by considering the QCD corrections
to the interference of the diagrams ${\cal M}_1$ and ${\cal M}_3$.
Consider first 
the correction due to the soft part of the diagram ${\cal M}_3$.
The operator which ``produces'' final hadronic 
state can be written as
\begin{equation}
\hat \ew \hat p \hat \ez \wm = -\hat p \left( 
\ew\cdot\ez+\frac {1}{2} [\hat \ew, \hat \ez] \right) \wm .
\label {10}
\end{equation}
Here we have used Eq.~(\ref {4}).
Note that because of the sum over polarizations of the $Z$ and $W$ bosons
there is no interference
between the first and the second structure in the above equation.
The amplitude ${\cal M}_1$ is  always proportional to the product
$\ew\cdot\ez$. 
Therefore in the soft correction only the first structure in eq.~(\ref{10})
contributes. The correction to it is again 
just the correction to the $W$ decay width.
As a result we find
\begin{equation}
\delta _4 ^{soft} = \frac {3}{4}.
\label {11}
\end{equation}

Now the ``hard'' part of the correction comes from the
interference of the $C_\mu $ structure of eq.~(\ref {9.5}) and eq.~(\ref {3}). 
Comparing coefficients and signs of the ``hard'' correction with the
corresponding Born one we conclude that the hard correction to
$H_4(x)$ is
\begin{equation}
\delta _4 ^{hard} = -~\frac {1}{2}
\label {12}
\end{equation}
and 
\begin{equation}
\delta _4 = \delta^{soft} _4 +\delta^{hard} _4 = 
\frac {1}{4}
\label {13}
\end{equation}
Clearly, the contribution of the diagram with the down-type quarks coupling 
to $W$--boson is the same up to trivial redefinitions of the coupling 
constant. 

In a similar manner we obtain the QCD correction to the square of the 
amplitudes ${\cal M}_2$ or ${\cal M}_3$ and for their interference.
The hard correction can be immediately 
determined from ${\cal M}^{hard}_{2,3}$.
For the soft part, however, one needs a full calculation. This is by no
means difficult and technically is equivalent to the calculation
of the QCD correction to the $e^+e^-$ annihilation to massless quarks.
All together this gives
\begin{equation}
\delta _1 = -~\frac {7}{12}.  
\label {14}
\end{equation}

We find also 
\begin{equation}
\delta _3 = -~\frac {5}{4}.
\end{equation}

Let us summarize the results of the QCD corrections to the quantities
$H_i(x)$:
\begin{equation}
\begin{array}{cccc}
\delta^{hard}_1 =-{8\over 3}, &
\delta^{hard}_2 = 0, &
\delta^{hard}_3 = -6, &
\delta^{hard}_4 =-{1\over 2}, 
\\
\delta^{soft}_1 ={25\over 12}, &
\delta^{soft}_2 ={3\over 4}, &
\delta^{soft}_3 ={19\over 4}, &
\delta^{soft}_4 ={3\over 4}. 
\end{array}
\label{qcdall}
\end{equation}
The division of the QCD corrections into the soft and hard parts helps
us to determine the proper energy scale of the strong coupling
constant.  From the explicit calculations which we have described it
is clear that the characteristic virtualities of the gluons are
$(M_Z-M_W)/2$ and $M_Z$ in the soft and hard parts, respectively.
Therefore, instead of eq.~(\ref{8.2}) we use
\be
H_i(x)=(1-x)^5 h_i^{(0)} \left( 1 
+ C_f \frac {\alpha _s\left(M_Z^2\right)}{\pi} \delta _i ^{hard}
+ C_f \frac {\alpha _s\left(
     \left[ {M_Z-M_W \over 2} \right]^2 \right) }{\pi} \delta _i ^{soft}
\right).
\ee

Among the four correction factors $\delta_i\equiv
\delta^{hard}_i+\delta^{soft}_i$  two can be compared with previously known
results.  $\delta_2$ is the well known QCD correction to (axial)vector
currents.  $\delta_1$, on the other hand, is the first term of the
expansion of the exact result for the square of the  diagram 1b (or
1c) which was recently found in a rather unusual way \cite{ck}.  The
complete mixed electroweak and QCD corrections to
the $Z$ boson decay width were computed in the limit of very small and
very large $W$ boson mass.  Their difference gives precisely the QCD
corrections to the emission of real $W$.  From the several known terms
of both expansions the general (very simple) 
formula for the coefficients was
guessed; the exact sum of this series is, however, rather complicated:
\be
\Gamma^{1b+1c}=
C_F {\alpha_s\over \pi} 
\Gamma_0
{g^{u2}_- + g^{d2}_-\over 2}
&& \hspace*{-4mm} \left[ -{7\over 4}
+{1\over 2}\ln x
+{3\over 32x}
- {9x\over 8} \ln x
+x^2 \left( {7\over 4} + {1\over 2} \ln x\right)
\right. \nonumber \\ && \left.
- {3x^3\over 32}
+ {2x^2\over 3}S(x) 
 - {2\over 3} S\left( {1\over x}\right)
\right].
\ee
Here
\be
S(x) &=& 
-{4\over x^2} \left(x^2+4x+1\right) \Li{4}
+{2\over x^2} \left(7x^2+16x+7\right) \Li{2}
\nonumber \\
&&-{30\over x^2} \left(1-x^2\right) \ln(1-x)
-{137\over 4} - {40\over x}
\nonumber \\
&& 
+\ln(x)\left[
{4\over x^2} \left(x^2+4x+1\right) \Li{3}
+{6\over x^2} \left(1-x^2\right) \Li{2}
-18 - {10\over x} \right]
\nonumber \\
&& 
+\ln^2(x)\left[
-{1\over x^2} \left(x^2+4x+1\right) \Li{2}
+{3\over x^2} \left(1-x^2\right) \ln(1-x)
+{23\over 4} + {4\over x} \right]
\ee
where $\Li{2,3,4}$ are polylogarithms \cite{lewin}.
An expansion of this formula around $x=1$ 
confirms our result for $\delta_1$.
One can expect that the exact result for the correction to the
interference of diagrams 1b and 1c would be even more complicated and
its full calculation is a rather daunting task.  Therefore it is
appropriate to use an expansion method described in this section.

\section {Conclusion}

We have presented a simple estimate of the ${\cal O}(\alpha _s)$ correction
to the decay rate $Z \to W^-u\bar d,W^+\bar u d$. 
Our approach is based on the expansion
of the event rate in powers of $(M_Z-M_W)/M_Z$.  
In principle, this expansion parameter is not very
small and the leading term in the expansion is not expected to
approximate the exact result with high accuracy.  However, it gives us
reasonable estimates of the lowest order and one--loop QCD corrections
to the decay width almost without effort.  Another important point
is that the accuracy of this approximation can be improved by
calculating further terms in the expansion, should need arise. Such a
calculation is by no means impossible.

Combining all QCD corrections presented in eq.~(\ref {qcdall}) we
arrive at the following correction to the decay rate:
\begin{equation}
\Gamma (Z \to W u \bar d) = \Gamma^{\rm Born}(Z \to W u \bar d) \left( 1 
+ {4\over 3} \frac{\alpha _s\left(M_Z^2\right)}{\pi} 
+ \frac {\alpha _s\left(
     \left[ {M_Z-M_W \over 2} \right]^2 \right) }{\pi}
\right)
\end{equation}
where we have used $C_F=4/3$.

For the values of the strong coupling constant we use 
$\alpha _s\left(M_Z^2\right)=0.12$  and 
$\alpha _s\left(
     \left[ {M_Z-M_W \over 2} \right]^2 \right)$ $=(0.25..0.30)$.
Numerically this gives the correction to the
decay rate $Z \to W u \bar d$ of the order of $13$--$15$ percent.

Let us have a closer look at the accuracy of our
formulas.  We can compare our approximate results 
(\ref {8}--\ref {8.2}) for the lowest order event rate with 
the results based on a complete  calculation \cite {MW}. We conclude
that our results based on the leading asymptotics to the event rate
give a $30$ percent accuracy. 
Note also that the accuracy with which the 
ratios $H_i/H_j$ are reproduced is even better and amounts to approximately
10\%.  We can therefore expect that the QCD corrections derived in 
this paper represent the complete unknown QCD corrections to the decay 
rate $Z \to Wud$ with the accuracy $10$--$20$ percent. 

Our final result is that the QCD corrections increase the decay rate
of the process $Z \to W u \bar d$ by approximately $14\pm 2$ percent.

\section*{Acknowledgments} 
We thank Professor Johann K\"uhn and 
Professor William Marciano for helpful comments.
This research was
supported by the grant BMBF 057KA92P and 
by the Graduierten\-kolleg ``Teilchenphysik''
of Mainz University.


\begin{thebibliography}{99}
\bibitem{MW} W.~Marciano, D.~Wyler, Z. Phys. {\bf C3} (1979) 181;
H.~Pietschmann, H.~Rupertsberg, K.~Svozil, Z. Phys. {\bf C12}
(1982) 367; 
H.~Rupertsberg, K.~Svozil, Acta Physica Austriaca, {\bf 54}
(1982) 255. 

\bibitem{LS} S.~C.~Lee and W.~C.~Su, Phys. Lett. {\bf B205} (1988) 569;
Z.~Phys. {\bf C40} (1988) 547.

\bibitem {Bij} 
E.~W.~N. Glover and J.~J. {van der Bij} (conv.),  
  in G. Altarelli, R. Kleiss, and C. Verzegnassi (eds.),
  {\em Z Physics at LEP 1},  CERN Yellow Report 89-08.

\bibitem{BK} E.~Braaten, A.~Kumar, Phys. Rev. {\bf D37} (1988) 3349. 

\bibitem{Alb}
D. Albert, W.J.~Marciano, Z.~Parsa, and
D.~Wyler, Nucl. Phys. {\bf B166}  (1980) 460.

\bibitem{ae}
S.G.~Gorishny, preprints JINR E2--86--176, E2--86--177 (Dubna 1986);
Nucl. Phys. {\bf B319} (1989) 633;
K.G.~Chetyrkin, Teor. Mat. Fiz. {\bf  75} (1988) 26; {\bf 76} (1988) 207;
V.A.~Smirnov,  Commun. Math. Phys. {\bf 134} (1990) 109;
K.G.~Chetyrkin, preprint MPI-PAE/PTh 13/91 (Munich, 1991).

\bibitem{ck}
A. Czarnecki and J.H.~K\"uhn, preprint TTP96-28, hep-ph/9608366.

\bibitem{lewin}
L. Lewin, {\em Polylogarithms and associated functions}
(North Holland, New York, 1981).

\end{thebibliography}
\end{document}